\documentclass[prd,showpacs]{revtex4}
\begin{document}

\title{On the Lagrangian description of dissipative systems}

\author{N.E. Mart\'{\i}nez-P\'erez}
 \email{nephtalieliceo@hotmail.com}
  \affiliation{Benem\'erita Universidad Aut\'onoma de Puebla, Facultad de Ciencias
F\'{\i}sico Matem\'aticas, P.O. Box 165, 72570 Puebla, M\'exico.}
 
\author{C. Ram\'{\i}rez}
 \email{cramirez@fcfm.buap.mx}
 \affiliation{Benem\'erita Universidad Aut\'onoma de Puebla, Facultad de Ciencias
F\'{\i}sico Matem\'aticas, P.O. Box 165, 72570 Puebla, M\'exico.}

 \date{February 16, 2018}

\begin{abstract}
We consider the Lagrangian formulation with duplicated variables of dissipative mechanical systems. The application of Noether theorem leads to physical observable quantities which are not conserved, like energy and angular momentum, and conserved quantities like the Hamiltonian, that generate symmetry transformations and do not correspond to observables. We show that there are simple relations among the equations satisfied by these two types of quantities. In the case of the damped harmonic oscillator, from the quantities obtained by Noether theorem follows the algebra of Feshbach and Tikochinsky. Further, if we consider the whole dynamics, the degrees of freedom separate into a physical and an unphysical sector. We analyze several cases, with linear and nonlinear dissipative forces; the physical consistency of the solutions is ensured observing that the unphysical sector has always the trivial solution.  
\end{abstract}

\maketitle

\section{Introduction}
The study of real physical systems requires the inclusion of external influences, whose origin is microscopic but frequently admit phenomenological descriptions, the damped harmonic oscillator is paradigmatic. The evolution of such systems is in general irreversible. Lagrangian formulations for these phenomenological theories are not straightforward \citep{bauer}, and there are multiple approaches in this direction. In \citet{dekker}, a review with 563 references, a presumably exhaustive analysis has been done, concluding to the date of this paper that none of the considered approaches leaded to a full satisfactory quantum formulation. In fact, a model independent description of dissipative systems is interesting for quite diverse fields like quantum optics \citep{walls}, quantum decoherence \citep{buchleitner}, general relativity \citep{gr1,gr2,cariglia}, string theory \citep{sen1,sen2}. Dissipative behavior appears also, in the form of information loss, within a proposal for a Planck scale deterministic approach for quantum mechanics in \citep{thooft1,thooft2}.  

The Lagrangian or Hamiltonian study of phenomenological approaches of dissipative systems has been done mainly by the introduction of additional variables \citep{bateman,schwinger}, by time dependent Lagrangians \citep{caldirola,caldirola1,caldirola2,caldirola3,schuch3,caldirola4}, and with complex actions \citep{complejo1,complejo2,complejo3,complejo4}. It has been addressed by coupling to heath baths \citep{heath1,heath2}, from which follow master equations \citep{dekker,master1,master2} and non-linear approaches \citep{nolineal1,nolineal2,nolineal3}. Recently a description by means of contact Hamiltonian mechanics has been proposed \citep{bravetti}. In \citep{bender} a generalized version has been studied, which models two coupled optical resonators, and which exploits the PT symmetry present in such systems \citep{benderpt}. 
These formulations and their quantization have been widely studied. However, despite this wide interest, this subject has still important open questions, see e.g. \citep{aldaya,aldaya1,chruscinski,chung}.

For conservative systems, the variational principle of Hamilton gives a way to obtain the equations of motion from an action, with the  physical trajectory determined by conditions in the past and in the future. From it can be obtained the Hamiltonian formalism and canonical quantization. The symmetries of the action lead to conservation laws, which can be obtained from Noether theorem, in particular for the energy, which coincides with the Hamiltonian. 
Dissipative forces lead to the violation of these conservation laws, making time dependent the otherwise conserved quantities. Thus, if we have a description with a time independent Lagrangian, like Bateman's one \citep{bateman}, the energy and the Hamiltonian will not be the same.
Further, the variation of paths beginning and ending at fixed points is not suitable. A proposal in this direction has been made by \citet{schwinger}, by the inclusion of a time reversed sector with a different dynamics, which corresponds to the doubling of variables or dual model of \citet{bateman}. Following Schwinger, the closed time-path formulation in quantum field theory has been developed, see e.g. \citep{chou}. In \citep{aldaya} it has been shown that the consideration of a Hamiltonian operator responsible for time evolution, along with an algebra with time dependent operators of position and momentum, leads to an operator algebra in terms of which the Hamiltonian corresponds to the Hamiltonian of Bateman. 
In a recent work \citep{galley1}, a generalization of the previous developments on doubling of variables has been proposed, with an action based on a conservative Lagrangian, and a generalized ``nonconservative potential'' which depends on both types of variables, along with a generalization of the Hamilton variational principle. In this approach the variation of the action is done with boundary conditions only at the initial time, independently for each of both variables, and at the final time these variables must coincide. A similar development for classical and quantum mechanics was given by means of an extension of the Closed Time Path formalism to classical mechanics by \citet{polonyi,polonyi2}, who in \citep{polonyi3} considers the issue of breaking of time reversal symmetry. 

The main interest in the study of phenomenological dissipative systems is on their quantum description. Classically, the doubled variable formalism allows to write the equations of motion and after that the additional variables are somehow discarded. In fact, these variables are considered as an artifice which takes account of the dissipative external influence, the whole system being isolated. However, from its construction, the nonconservative Lagrangian has not the standard form due to the time reversed characteristics of the additional sector, i.e. the kinetic term is not positive definite and the potential appears with an unstable term. Thus, an interpretation of its outcome as a whole is not obvious. On the other side, in a quantum theory every interacting degree of freedom in general contributes to the probabilities, spectra and mean values, as they form part of the operator algebra. Thus, it would be desirable to consider the classical theory taking into account the whole dynamics. Moreover, a general knowledge of the relevant quantities in the theory, as delivered e.g. by Noether theorem, is necessary for the definition of the Hilbert space. Actually, in the doubled variable approach, Noether theorem has been applied considering the conservation laws of the conservative part, and these laws are violated due to the dissipative terms \citep{galley2,polonyi2}. Furthermore, Noether theorem has been applied in similar approaches to the symmetries of the whole doubled variables action in \citep{mendes,nos}, and for time dependent lagrangians in \citep{cervero}. Otherwise the symmetries can be studied considering the operator algebra of the system, as done in \citep{aldaya,aldaya1,cariglia}

In this paper, we start from the nonconservative Lagrangian of Galley. As for our considerations we use only the Euler-Lagrange equations, we do not apply at this stage the final time boundary conditions required for the variation. We give a formulation for Noether theorem, considering the transformations which let invariant only the conservative Lagrangian, as well as the transformations which let invariant the whole nonconservative Lagrangian. Among the last, there may be transformations that mix both types of variables \citep{lukierski,mendes}. The application of Noether theorem for the symmetries of the nonconservative Lagrangian leads to conserved quantities that are generators of the corresponding transformations. On the other side, the application of Noether theorem to the symmetries of the conservative Lagrangian leads to the violation of the conservation laws of the corresponding physical quantities, energy, angular momentum, etc., which appear in two copies each one, due to the doubling of variables. It follows that these quantities, conserved and non conserved, are not independent. For example the conservation equation of the Hamiltonian follows from the equations satisfied by the energy of the original conserved system and the energy of the doubled system.
Further, we consider the dynamics for all degrees of freedom, in standard mechanical terms, considering that in a quantum theory the fluctuations of all variables must be taken into account. This means that the equations of motion of all the variables, including the doubled ones, should be solved. As turns out from the general form of the equations of motion, and from the analysis of examples, there are physical and unphysical solutions, where the former have the expected behavior resulting from dissipation, i.e. decreasing velocity and energy, opposite to the second case, whose velocity and energy in general increase steadily. Thus there are two sectors, in general corresponding to these types of solutions. Although the equations of motion of both sectors are in general coupled, the unphysical sector has always the trivial, vanishing solution, and it must be taken in absence of any other consistent solution. Such an argument has been used by Dirac in \cite{dirac}.
This result is consistent with the variational principle, which restricts the trajectories in the unphysical sector so that at the final time these variables and their first derivatives coincide. If we consider this restriction for the solutions, only the mentioned trivial solution of the unphysical sector satisfies it. This leads to the ``physical limit'' of \citet{galley1}. 

In Section II we give short review of the formulation of Galley. In section III we consider the Hamiltonian formulation. In Section IV we work out the Noether theorem and discuss its consequences. In Section V we discuss the examples of free particle, free fall, harmonic oscillator and central forces for linear dissipation, and in Section VI we consider nonlinear dissipation. In the last Section we draw some conclusions.

\section{Lagrangian Formulation}
The Lagrangian formulation of dissipative systems has as one of its paradigms the Bateman formulation for the damped harmonic oscillator \cite{bateman}, with Lagrangian
\begin{equation}
L=\frac{1}{2}\left[\dot{x}\dot{y}-\kappa xy+\gamma(x\dot{y}-y\dot{x})\right].\label{bateman}
\end{equation}
The Euler-Lagrange equation of this action for  the variable $x$, for $\gamma>0$, describes the damped harmonic oscillator, and $y$ is an auxiliary degree of freedom associated with the environment, its dynamics is discarded. 
The kinetic term of (\ref{bateman}) is diagonalized by the transformation $q_1=x+y$, $q_2=x-y$, i.e. 
$L=\frac{1}{2}(\dot{q}_1^2-\kappa q_1^2)-\frac{1}{2}(\dot{q}_2^2-\kappa q_2^2)+\gamma(q_2\dot{q}_1-q_1\dot{q}_2)$. 
This Lagrangian is antisymmetric under the interchange $q_1\leftrightarrow q_2$, and it inspired generalized formulations by \citet{galley1} and \citet{polonyi}, with a conservative Lagrangian $L(q,\dot{q})$ as starting point, whose degrees of freedom $q$, in general $n$-dimensional, are doubled $q\rightarrow(q_1,q_2)$ in order to write an action
\begin{equation}\label{accion0}
S=\int_{t_i}^{t_f}L({q}_1,{\dot{q}}_1)dt-\int_{t_i}^{t_f}L({q}_2,{\dot{q}}_2)dt=\int_{t_i}^{t_f}L({q}_1,{\dot{q}}_1)dt+\int_{t_f}^{t_i}L({q}_2,{\dot{q}}_2)dt.
\end{equation}
To this action is added the nonconservative potential or influence functional $K(q_1,\dot q_1,q_2,\dot q_2)$, which depends on both variables and is antisymmetric under the interchange $1\leftrightarrow 2$, i.e.
\begin{equation}
K(q_1,\dot q_1,q_2,\dot q_2)=-K(q_2,\dot q_2,q_1,\dot q_1).\label{antisym}
\end{equation}
Thus the generalized action is
\begin{equation}
S=\int_{t_i}^{t_f}\Lambda(q_1,\dot q_1,q_2,\dot q_2)dt=\int_{t_i}^{t_f}\left[L({q}_1,{\dot{q}}_1)-L({q}_2,{\dot{q}}_2)+K(q_1,\dot q_1,q_2,\dot q_2)\right]dt.\label{ncl}
\end{equation}
The variation can be done by the usual Hamilton's principle, with fixed variables at initial and final times, leading to the Euler-Lagrange equations. However \citep{galley1,polonyi}, in this case the effective interaction obtained for an environment would be reversible, as a consequence of the time-symmetric boundary conditions of the variational principle. These conditions lead also to causality problems, as dissipative processes are determined by initial conditions. 
These drawbacks are overcome by a modification of the variational conditions at final time. It is striking that this feature can be incorporated into a variational principle for (\ref{ncl}), consistently with the Euler-Lagrange equations. In fact, action (\ref{accion0}) corresponds to a variable $q_{2}(t)$ running back in time, as in the closed path-time (CPT) approach  \citep{schwinger,chou}, and its variation is used to control the variation of $q_1(t)$ at the final time by the coupling $q_{1}(t_f)=q_2(t_f)$. In \citep{polonyi} both variables are arranged as one, beginning at $t_i$, and finishing at $2t_f-t_i$.  Thus, the boundary conditions for the variation are that, at the initial time both variables are independently fixed and their variations vanish, and at the final time they coincide, with an otherwise arbitrary variation. That is, 
\begin{equation}
\delta{q}_1(t_i)=\delta{q}_2(t_i)=0, \quad {q}_1(t_f)={q}_2(t_f)\ \ {\rm and }\ \ {\dot {q}}_1(t_f)={\dot {q}}_2(t_f).\label{cond}
\end{equation}
Hence for the variations at the final time the only condition is $\delta{q}_1(t_f)=\delta{q}_2(t_f)$.
Actually, this variation contains the usual variation, hence in any case leads to the usual Euler-Lagrange equations \citep{brunt}.
Thus, if we denote $L_1=L(q_1,\dot{q}_1)$ and  $L_2=L(q_2,\dot{q}_2)$, the variation gives
\begin{eqnarray}
\delta S&=&\int_{t_{i}}^{t_{f}}\delta\Lambda(q_1,\dot q_1,q_2,\dot q_2)dt
=\left.\left[\delta q_{1}\left(\frac{\partial L_1}{\partial{\dot q}_{1}}+\frac{\partial K}{\partial{\dot q}_{1}}\right)+\delta q_{2}\left(-\frac{\partial L_2}{\partial{\dot q}_{2}}+\frac{\partial K}{\partial{\dot q}_{2}}\right)\right]\right|_{t=t_{f}}\nonumber\\
&+&\int_{t_{i}}^{t_{f}}\left[\delta q_{1}\left(\frac{\partial\Lambda}{\partial q_{1}}-\frac{d}{dt}\frac{\partial\Lambda}{\partial{\dot q}_{1}}\right)+\delta q_{2}\left(\frac{\partial\Lambda}{\partial q_{2}}-\frac{d}{dt}\frac{\partial\Lambda}{\partial{\dot q}_{2}}\right)\right]dt.
\end{eqnarray}
The boundary terms vanish after taking into account the boundary conditions and the antisymmetry of $K$, from which follows
\begin{equation}
\left.\left(\frac{\partial K}{\partial{\dot q}_{1}}+\frac{\partial K}{\partial{\dot q}_{2}}\right)\right|_{q_1=q_2,\ \dot q_1=\dot q_2}=0.\label{derk}
\end{equation}
Thus, the equations of motion are 
\begin{eqnarray}
\frac{\partial\Lambda}{\partial q_{1}}-\frac{d}{dt}\frac{\partial\Lambda}{\partial{\dot q}_{1}}&=&0,\quad {\rm and}\label{eq1}\\
\frac{\partial\Lambda}{\partial q_{2}}-\frac{d}{dt}\frac{\partial\Lambda}{\partial{\dot q}_{2}}&=&0,\label{eq2}
\end{eqnarray}
which can be written as
\begin{eqnarray}
\left(\frac{\partial}{\partial q_{1}}-\frac{d}{dt}\frac{\partial}{\partial{\dot q}_{1}}\right)L(q_1,\dot{q_1})&=&-(F_K)_1,\quad {\rm and}\label{eqf1}\\
\left(\frac{\partial}{\partial q_{2}}-\frac{d}{dt}\frac{\partial}{\partial{\dot q}_{2}}\right)L(q_2,\dot{q_2})&=&(F_K)_2,\label{eqf2}
\end{eqnarray}
where $(F_K)_1=\left(\frac{\partial}{\partial q_{1}}-\frac{d}{dt}\frac{\partial}{\partial{\dot q}_{1}}\right)K(q_1,\dot{q_1},q_2,\dot{q_2})$ and $(F_K)_2=\left(\frac{\partial}{\partial q_{2}}-\frac{d}{dt}\frac{\partial}{\partial{\dot q}_{2}}\right)K(q_1,\dot{q_1},q_2,\dot{q_2})$
are the nonconservative forces. 

In terms of Bateman's or ``light cone'' variables, which we will call from now on $(q_{+},q_{-})$, and which are related to the ``cartesian'' variables by $q_{\pm}=\frac{1}{2}(q_1\pm q_2)$, the boundary conditions are 
\begin{equation}
\delta q_{\pm}(t_i)=0, \quad q_{-}(t_f)=0\quad \textnormal{and}\quad \dot q_{-}(t_f)=0, \label{condv}
\end{equation}
and the equations of motion are 
\begin{equation}
\left(\frac{\partial}{\partial q_{\pm}}-\frac{d}{dt}\frac{\partial}{\partial{\dot q}_{\pm}}\right)\Lambda(q_{+},q_{-},\dot{q}_{+},\dot{q}_{-})=0,\label{eqpm}
\end{equation}
i.e.
\begin{eqnarray}
\left(\frac{\partial}{\partial q_{-}}-\frac{d}{dt}\frac{\partial}{\partial{\dot q}_{-}}\right)L_{-}&=&-(F_K)_{-} \quad {\rm and}\label{eqp}\\
\left(\frac{\partial}{\partial q_{+}}-\frac{d}{dt}\frac{\partial}{\partial{\dot q}_{+}}\right)L_{-}&=&-(F_K)_{+},\label{eqm}
\end{eqnarray} 
where $L_{-}=L(q_1,\dot{q}_1)-L(q_2,\dot{q}_2)=L(q_{+}+q_{-},\dot q_{+}+\dot q_{-})- L(q_{+}-q_{-},\dot q_{+}-\dot q_{-})$, and $(F_K)_{\pm}=\left(\frac{\partial}{\partial q_{\pm}}-\frac{d}{dt}\frac{\partial}{\partial{\dot q}_{\pm}}\right)K$. 
Note that $q_1\leftrightarrow q_2$ implies $q_{\pm}\leftrightarrow\pm q_{\pm}$, $L_{-}\leftrightarrow -L_{-}$ and $K\leftrightarrow -K$, hence $\Lambda\leftrightarrow -\Lambda$. This antisymmetry has the important consequence that $\Lambda$, $L_{-}$, $K$, and their derivatives with respect to $q_{+}$ and $\dot q_{+}$, vanish identically when $q_1(t)=q_2(t)$, i.e. $q_{-}(t)=0$. This means that equation (\ref{eqm}) has always the trivial solution $q_{-}(t)=0$. Moreover, in this case, the momentum $p_{-}(t)=\frac{\partial\Lambda}{\partial \dot{q}_{+}}$ vanishes as well. Note that the equations of motion are the same if a total derivative $\frac{d}{dt}f(q_1,q_2)$ is added to the Lagrangian in (\ref{ncl}), where $f(q_1,q_2)$ is antisymmetric.  

The preceding approach has been applied mainly to: a) Derivation of effective actions by the inclusion of environmental variables, e.g. harmonic oscillators \citep{galley1,polonyi}, in this case no nonconservative potential is required; b) Systems subject to dissipative forces like the damped harmonic oscillator, where the starting point is a conservative system, with a nonconservative potential $K(q_1,\dot q_1,q_2,\dot q_2)$. In both cases the boundary conditions at $t_f$ in (\ref{condv}) are applied on the solutions of the equations of motion. In the case a) the boundary conditions contribute to the causal consistency of the effective action, and the physically relevant degree of freedom is $q_{+}(t)$. In the case b) the boundary conditions lead to the trivial solution $q_{-}(t)=0$, which amounts to the physical limit of \citet{galley1}. In the present work we will consider only cases of type b). However, if we are interested on the quantum theory, it is meaningless to set $q_{-}(t)=0$. Thus, although in this paper we consider only classical theory, we will not impose these conditions. We will rather study the consequences of the Euler-Lagrange equations (\ref{eqpm}). It turns out that, as can be guessed already from Bateman equations of motion, the solutions of the equations of motion for $q_{-}$ turn out to be physically meaningless, unless the trivial solution is taken. We show explicitly that in many relevant cases $q_{-}$ has always this behaviour. 

The formalism of this section can be straightforwardly generalized for any number of degrees of freedom, and for any conservative Lagrangian. 

Note that in the conservative limit, i.e. if the nonconservative potential $K$ is set to zero, equations (\ref{eq1}) and (\ref{eq2}) describe two identical independent copies of the conservative system. In fact, in this case equations (\ref{eqp}) and (\ref{eqm}) decouple by the transformation $(q_{+},q_{-})\rightarrow(q_1,q_2)$. 
\section{Hamiltonian formulation}
If the conservative Lagrangian is $L(q,\dot q)$, then its canonical momenta are $p=\partial L/\partial\dot q$ and its Hamiltonian $H(q,p)=\dot qp-L(q,\dot q)$. Thus for the nonconservative Lagrangian, for consistency with the conservative sector, the convention is that the momenta are \citep{galley2}
\begin{eqnarray}
p_1&=&\frac{\partial\Lambda}{\partial\dot q_1}=\frac{\partial}{\partial\dot q_1}[L(q_1,\dot q_1)+K(q_1,\dot q_1,q_2,\dot q_2)],\label{p1}\\
p_2&=&-\frac{\partial\Lambda}{\partial\dot q_2}=\frac{\partial}{\partial\dot q_2}[L(q_2,\dot q_2)-K(q_1,\dot q_1,q_2,\dot q_2)].\label{p2}
\end{eqnarray}
Hence the Hamiltonian is
\begin{equation}
H(q_1,p_1,q_2,p_2)=\dot q_1p_1-\dot q_2p_2-\Lambda(q_1,\dot q_1,q_2,\dot q_2)=2(\dot q_{+}p_{-}-\dot q_{-}p_{+})-\Lambda.\label{hnc}
\end{equation}
This system is regular if $\det(\partial p_a/\partial\dot q_b)\neq0$ $(a,b=1,2)$, i.e.
\begin{equation}
\det\left(
\begin{array}{cc}
\frac{\partial^2[L(q_1,\dot{q}_1)+K]}{\partial \dot{q}_1^2}&\frac{\partial^2K}{\partial \dot{q}_1\partial \dot{q}_2}\\\frac{\partial^2K}{\partial \dot{q}_1\partial \dot{q}_2}&\frac{\partial^2[L(q_2,\dot{q}_2)-K]}{\partial \dot{q}_2^2}
\end{array}
\right)\neq0.
\end{equation}
In this case, the solutions of the system (\ref{p1}), (\ref{p2}) are in general of the form $\dot q_1=\dot q_1(q_1,p_1,q_2,p_2)$ and $\dot q_2=\dot q_2(q_1,p_1,q_2,p_2)$.
Thus, the equations of motion are 
\begin{eqnarray}
\dot q_1=\frac{\partial H}{\partial p_1},\qquad \dot q_2=-\frac{\partial H}{\partial p_2},\label{heq1}\\ 
\dot p_1=-\frac{\partial H}{\partial q_1},\qquad \dot p_2=\frac{\partial H}{\partial q_2},\label{heq2}
\end{eqnarray}
which are equivalent to (\ref{eq1}) and (\ref{eq2}). 
Thus, the evolution is given by $\dot f(q_1,p_1,q_2,p_2)=\left\{f,H\right\}$, where the generalized Poisson brackets are
\begin{equation}
\left\{f,g\right\}\equiv\frac{\partial f}{\partial q_1}\frac{\partial g}{\partial p_1}-\frac{\partial f}{\partial p_1}\frac{\partial g}{\partial q_1}-\left(\frac{\partial f}{\partial q_2}\frac{\partial g}{\partial p_2}-\frac{\partial f}{\partial p_2}\frac{\partial g}{\partial q_2}\right).\label{poisson}
\end{equation}
Therefore, the nonvanishing Poisson brackets among canonical variables are
\begin{equation}
\{q_1,p_1\}=1,\qquad \{q_2,p_2\}=-1,\label{cpc}
\end{equation}
or, in light cone coordinates
\begin{equation}
\{q_{+},p_{-}\}=\frac{1}{2},\qquad \{q_{-},p_{+}\}=\frac{1}{2},\label{cpcl}
\end{equation}
where 
\begin{equation}
p_{\pm}=\frac{1}{2}\frac{\partial\Lambda}{\partial q_{\mp}}.\label{ppm}
\end{equation}
It is obvious that the transformations of the form $Q_1=Q_1(q_1,p_1)$, $P_1=P_1(q_1,p_1)$, $Q_2=Q_2(q_2,p_2)$ and $P_2=P_2(q_2,p_2)$, which preserve the form of the equations (\ref{heq1}) and (\ref{heq2}), are canonical transformations in the usual sense.
\section{Noether theorem} \label{sec:noether}
A characteristics of the Lagrangian descriptions of nonconservative systems, is that the invariances of the equations of motion and of the Lagrangian might not coincide \citep{sarlet}, as can happen when the equations of motion differ from the Euler-Lagrange equations by a nonconstant factor. In the present case, the construction of the nonconservative Lagrangian from the conservative Lagrangian plus the nonconservative potential, allows a coincidence of these invariances. Hence the implementation of Noether theorem seems to be meaningful.
In \citet{galley2}, starting from the Noether theorem for the conservative system and the nonconservative equations of motion, the violation of the corresponding conservation laws is derived for the nonconservative system.

In the usual case, the Noether theorem can be formulated from a variation of the Lagrangian, as the boundary conditions play no role. If we transform the Lagrangian $L(q,\dot q,t)$ under a time translation $t\rightarrow t+\delta t$ and internal transformations $\delta_\alpha q$, then
\begin{eqnarray}
\delta L&=&\delta t\left(\dot q\frac{\partial L}{\partial q}+\ddot q\frac{\partial L}{\partial\dot q}+\frac{\partial L}{\partial t}\right)+\delta_\alpha q\frac{\partial L}{\partial q}+\delta_\alpha\dot q\frac{\partial L}{\partial\dot q}\nonumber\\
&=&\frac{d}{dt}\left[(\delta t\dot q+\delta_\alpha q)\frac{\partial L}{\partial\dot q}\right]+(\delta t\dot q+\delta_\alpha q)\left(\frac{\partial L}{\partial q}-\frac{d}{dt}\frac{\partial L}{\partial\dot q}\right)+\delta t\frac{\partial L}{\partial t}.\label{deltaL}
\end{eqnarray}
Further, equating the right hand side of this equation with $\delta t\frac{dL}{dt}+\delta_\alpha L$ gives
\begin{equation}
\frac{d}{dt}\left[\delta t\left(\dot q\frac{\partial L}{\partial\dot q}-L\right)+\delta_\alpha q\frac{\partial L}{\partial\dot q}\right]
=-(\delta t\dot q+\delta_\alpha q)\left(\frac{\partial L}{\partial q}-\frac{d}{dt}\frac{\partial L}{\partial\dot q}\right)-\delta t\frac{\partial L}{\partial t}+\delta_\alpha L,\label{noether}
\end{equation}
from which, taking into account the equations of motion and the invariance of the action, Noether theorem follows. In other words, for solutions of the equations of motion, the Hamiltonian and the internal charges satisfy $\frac{dH}{dt}=\frac{d}{dt}\left(\dot q\frac{\partial L}{\partial\dot q}-L\right)=-\frac{\partial L}{\partial t}$ and $\frac{dJ_\alpha}{dt}=\frac{d}{dt}\left(\delta_\alpha q\frac{\partial L}{\partial\dot q}\right)=\delta_\alpha L$. Thus, if the Lagrangian does not depend explicitly on time and is invariant under the internal transformations, $H$ and $J$ are conserved quantities. Note that a somewhat different conservation law will follow if $\delta_\alpha L$ is a total time derivative.

This result can be applied to the nonconservative action $\Lambda(q_1,\dot q_1,q_2,\dot q_2)$, taking into account the equations of motion (\ref{eq1}) and (\ref{eq2}). In this case, the Hamiltonian and nonconservative currents satisfy
\begin{eqnarray}
\frac{dH}{dt}\equiv\frac{d}{dt}\left(\dot q_1\frac{\partial \Lambda}{\partial\dot q_1}+\dot q_2\frac{\partial \Lambda}{\partial\dot q_2}-\Lambda\right)=-\frac{\partial\Lambda}{\partial t},\label{ham}\\
\frac{d{\cal J}_\alpha}{dt}\equiv\frac{d}{dt}\left(\delta_\alpha q_1\frac{\partial\Lambda}{\partial\dot q_1}+\delta_\alpha q_2\frac{\partial\Lambda}{\partial\dot q_2}\right)=\delta_\alpha\Lambda,\label{jota}\\
\frac{d\tilde{\cal J}_\beta}{dt}\equiv\frac{d}{dt}\left(\tilde\delta_\beta q_1\frac{\partial\Lambda}{\partial\dot q_1}+\tilde\delta_\beta q_2\frac{\partial\Lambda}{\partial\dot q_2}\right)=\tilde\delta_\beta\Lambda,\label{jotamix}
\end{eqnarray}
where $\delta_\alpha q$ are internal transformations which do not mix $q_1$ and $q_2$, and $\tilde\delta_\beta q$ are transformations which mix $q_1$ and $q_2$.

Further, writing (\ref{noether}) separatedly for $L_1\equiv L(q_1,\dot{q}_1)$ and for $L_2\equiv L(q_2,\dot{q}_2)$, and taking into account the equations of motion (\ref{eqf1}) and (\ref{eqf2}), leads to
\begin{eqnarray}
\frac{d}{dt}\left[\delta t\left(\dot q_1\frac{\partial L_1}{\partial\dot q_1}-L_1\right)+\delta_\alpha q_1\frac{\partial L_1}{\partial\dot q_1}\right]=(\delta t\dot q_1+\delta_\alpha q_1)(F_K)_1-\delta t\frac{\partial L_1}{\partial t}+\delta_\alpha L_1,\label{deltaL1}\\
\frac{d}{dt}\left[\delta t\left(\dot q_2\frac{\partial L_2}{\partial\dot q_2}-L_2\right)+\delta_\alpha q_2\frac{\partial L_2}{\partial\dot q_2}\right]=-(\delta t\dot q_2+\delta_\alpha q_2)(F_K)_2-\delta t\frac{\partial L_2}{\partial t}+\delta_\alpha L_2,\label{deltaL2}
\end{eqnarray}
from which follow
\begin{eqnarray}
\frac{dE_1}{dt}&\equiv&\frac{d}{dt}\left(\dot q_1\frac{\partial L_1}{\partial\dot q_1}-L_1\right)
=\dot q_{1}(F_K)_1-\frac{\partial L_1}{\partial t},\label{noncec3}\\
\frac{dE_2}{dt}&\equiv&\frac{d}{dt}\left(\dot q_2\frac{\partial L_2}{\partial\dot q_2}-L_2\right)
=-\dot q_{2}(F_K)_2-\frac{\partial L_2}{\partial t},\label{noncec4}\\
\frac{dJ_{\alpha1}}{dt}&\equiv&\frac{d}{dt}\left(\delta_\alpha q_1\frac{\partial L_1}{\partial\dot q_1}\right)
=\delta_\alpha q_{1}(F_K)_1+\delta_\alpha L_1,\label{jota2}\\
\frac{dJ_{\alpha2}}{dt}&\equiv&\frac{d}{dt}\left(\delta_\alpha q_2\frac{\partial L_2}{\partial\dot q_2}\right)
=-\delta_\alpha q_{2}(F_K)_2+\delta_\alpha L_2.\label{jota3}
\end{eqnarray}
It turns out that from these equations follow (\ref{ham}) and (\ref{jota}), as well as equations for the quantities $E=\frac{1}{2}(E_1+E_2)$ and $J_\alpha=\frac{1}{2}(J_{\alpha1}+J_{\alpha2})$. Obviously, if we add (\ref{noncec3}) and (\ref{noncec4}), and then (\ref{jota2}) and (\ref{jota3}), we get
\begin{eqnarray}
2\frac{dE}{dt}&=&\frac{d}{dt}\left[\dot q_1\frac{\partial L(q_1)}{\partial\dot q_1}+\dot q_2\frac{\partial L(q_2)}{\partial\dot q_2}-L(q_1)-L(q_2)\right]\nonumber\\
&=&\dot q_{1}(F_K)_1-\dot q_{2}(F_K)_2-\frac{\partial(L_1+L_2)}{\partial t},\label{noncec1}\\
2\frac{dJ_\alpha}{dt}&=&\frac{d}{dt}\left[\delta_\alpha q_1\frac{\partial L(q_1)}{\partial\dot q_1}+\delta_\alpha q_2\frac{\partial L(q_2)}{\partial\dot q_2}\right]\nonumber\\
&=&\delta_\alpha q_{1}(F_K)_1-\delta_\alpha q_{2}(F_K)_2+\delta_\alpha(L_1+L_2).\label{noncec2}
\end{eqnarray}
However, for (\ref{ham}) and (\ref{jota}) is not as simple. We obtain (\ref{ham}) by subtracting (\ref{noncec4}) from (\ref{noncec3}), and take into account the identity 
\begin{eqnarray}
\dot q_{1}(F_K)_1+\dot q_{2}(F_K)_2\equiv-\frac{d}{dt}\left(\dot q_{1}\frac{\partial K}{\partial\dot q_1}+\dot q_{2}\frac{\partial K}{\partial\dot q_2}-K\right)-\frac{\partial K}{\partial t}.\label{fkk}
\end{eqnarray}
Further, (\ref{jota}) is obtained subtracting (\ref{jota3}) from (\ref{jota2}) and considering the identity
\begin{eqnarray}
\delta_\alpha q_{1}(F_K)_1+\delta_\alpha q_{2}(F_K)_2\equiv-\frac{d}{dt}\left(\delta_\alpha q_{1}\frac{\partial K}{\partial\dot q_1}+\delta_\alpha q_{2}\frac{\partial K}{\partial\dot q_2}\right)+\delta_\alpha K.
\end{eqnarray}
From equation (\ref{fkk}) follows in particular, that if $K$ is first-degree homogeneous in the velocities and time independent, as happens for the damped harmonic oscillator, the right hand side of (\ref{fkk}) vanishes, and 
\begin{equation}
H=E_1-E_2,\label{ham12}
\end{equation}
which in \citep{cariglia} is identifies with the total energy, as it is constant and the system is closed.

In conclusion, for the symmetries of the conservative system, equations (\ref{noncec3})-(\ref{jota3}) give the violation of the conservation of the energies and of the charges of the internal symmetries, which otherwise are conserved in the absence of dissipation. Further, equations (\ref{noncec3})-(\ref{jota3}) are equivalent to (\ref{ham}), (\ref{jota}), (\ref{noncec1}) and (\ref{noncec2}), from which follow the violation of the conservation of $E$ (\ref{noncec1}), and $J$ (\ref{noncec2}). Moreover, for the Hamiltonian and the nonconservative charges ${\cal J_\alpha}$, follow $\frac{dH}{dt}=-\frac{\partial K}{\partial t}$ and $\frac{d{\cal J}_\alpha}{dt}=\delta_\alpha K$; hence if $K$ does not depend explicitly on time and is invariant under internal transformations, these quantities are conserved. 

From equation (\ref{jotamix}) follows that for symmetries of the nonconservative Lagrangian that mix $q_1$ and $q_2$, there are conserved quantities $\tilde{\cal J}_\beta$, which do not have correspondence for the conservative system. The quantities $H$, ${\cal J}$ and $\tilde{\cal J}$ generate the corresponding transformations of the variables of the doubled system, i.e. time translations, internal transformations of the conservative system, and transformations which mix $q_1$ and $q_2$.

All the computations of this section can be straightforwardly generalized for any number of degrees of freedom. 

Note that in Galley's physical limit, (\ref{ham}) and (\ref{jota}) vanish identically due to the antisymmetry of $\Lambda$, and (\ref{noncec1}) coincides with the corresponding result obtained in \citep{galley1} for a time independent conservative Lagrangian.  

\section{Examples}
In this section we discuss some examples, which in general have a physical sector with dissipation, and an unphysical sector, with unbounded increasing energy for nontrivial solutions. These two sectors are described in light cone coordinates, $q_{+}$ for the physical sector, and $q_{-}$ for the unphysical one. As remarked after equation (\ref{eqmv2}), the equations of motion for the unphysical sector have always the trivial solution $q_{-}(t)=0$. 

\subsection{Linear dissipative forces}
Let us first consider an action with an arbitrary conservative potential, $L=\frac{m}{2}\dot{q}^2-V(q)$, and a nonconservative potential which corresponds to a force linear in velocity, acting opposite to it \citep{galley1}
\begin{equation}
K(q_{1},\dot q_{1},q_{2},\dot q_{2})=-\frac{c}{2}(q_{1}\dot q_{2}-q_{2}\dot q_{1})
=-c\left(q_{-}\dot{q}_{+}-q_{+}\dot{q}_{-}\right),\label{kgen}
\end{equation}
where $q$ can be an $n$-dimensional vector and the products $SO(n)$ invariant. In the following we will consider $n=1$, unless otherwise stated. 
Hence the nonconservative Lagrangian is
\begin{eqnarray}
\Lambda(q_1,q_2,\dot q_1,\dot q_2)=\frac{m}{2}(\dot q_1^{2}-\dot q_2^{2})-V(q_1)+V(q_2)-\frac{c}{2}(q_{1}\dot q_{2}-q_{2}\dot q_{1})\label{lambdagral}\nonumber\\
=2m\dot{q}_{+}\dot{q}_{-}-V(q_{+}+q_{-})+V(q_{+}-q_{-})-c\left(q_{-}\dot{q}_{+}-q_{+}\dot{q}_{-}\right). 
\end{eqnarray}
This Lagrangian is invariant under time translations and depending on the form of the potential $V(q)$, it could have other symmetries. In particular, the kinetic term and the nonconservative potential have an $SO(1,1)$ symmetry  \citep{mendes}
\begin{equation}
\delta q_1=\eta q_2,\qquad \delta q_2=\eta q_1,\label{lorentz}
\end{equation}
i.e. $q_{\pm}'=e^{\pm\eta}q_{\pm}$.
If the potential $V(q)$ is invariant under translations $\delta q_1=a_1$ and $=\delta q_2=a_2$, the Lagrangian transforms by a total derivative. 
The canonical momenta (\ref{p1}) and (\ref{p2}) are $p_1=m\dot q_{1}+\frac{c}{2}q_2$ and $p_2=m\dot q_{2}+\frac{c}{2}q_1$ or $p_{\pm}=m\dot{q}_{\pm}\pm\frac{c}{2}q_{\pm}$. 
The Hamiltonian is $H=\frac{1}{2m}\left(p_1-\frac{c}{2}q_2\right)^2-\frac{1}{2m}\left(p_2-\frac{c}{2}q_1\right)^2+V(q_1)-V(q_2)$, and can be written also as
\begin{equation}
H=\frac{2}{m}\left(p_{+}-\frac{c}{2}q_{+}\right)\left(p_{-}+\frac{c}{2}q_{-}\right)+V(q_{+}+q_{-})-V(q_{+}-q_{-}). \label{hamg}
\end{equation}
The nonconservative forces are $(F_K)_1=-c\dot{q}_2$ and $(F_K)_2=c\dot{q}_1$. From Noether theorem the energies $E_1=\frac{m}{2}\dot{q}_1^2+V(q_1)$ and $E_2=\frac{m}{2}\dot{q}_2^2+V(q_2)$ satisfy $\frac{dE_1}{dt}=\frac{dE_2}{dt}=-c\dot{q}_1\dot{q}_2$, consistently with the Hamiltonian given by (\ref{ham12}). The sum of these energies is $E=\frac{1}{2}(E_{1}+E_{2})=E_{+}+E_{-}+\frac{1}{2}\left[V(q_1)+V(q_2)\right]$, where
\begin{equation}
E_{\pm}=\frac{1}{2m}\left(p_{\pm}\mp\frac{c}{2} q_{\pm}\right)^2.\label{epm}
\end{equation}

If the potential is invariant under translations, then from (\ref{jota2}) and (\ref{jota3}), the momenta of the conservative theory $P_1=m\dot{q}_1=p_1-\frac{c}{2}q_2$ and $P_2=m\dot{q}_2=p_2-\frac{c}{2}q_1$ satisfy $\frac{dP_1}{dt}=-c\dot q_2$ and $\frac{dP_2}{dt}=-c\dot q_1$; hence $\frac{d}{dt}(p_1+\frac{c}{2}q_2)=0$ and $\frac{d}{dt}(p_2+\frac{c}{2}q_1)=0$. The last conserved quantities generate the phase space translations $\delta q_1=a_1$, $\delta q_2=a_2$, $\delta p_1=\frac{c}{2}a_2$, and $\delta p_2=\frac{c}{2}a_1$. Subtracting the equations satisfied by $P_1$ and $P_2$, we get the conservation equation for the generator of translations 
\begin{equation}
P=P_1-P_2-c(q_1-q_2)=p_1-p_2-\frac{c}{2}(q_1-q_2).\label{trans}
\end{equation}
The equations of motion are
\begin{eqnarray}
m\ddot{q}_{+}+2c\dot{q}_{+}-\frac{\partial}{\partial q_{-}}[V(q_{+}+q_{-})-V(q_{+}-q_{-})]&=&0,\label{eqmv1}\\
m\ddot{q}_{-}-2c\dot{q}_{-}-\frac{\partial}{\partial q_{+}}[V(q_{+}+q_{-})-V(q_{+}-q_{-})]&=&0.\label{eqmv2}
\end{eqnarray}
The second equation contains a force with the opposite sign as the first equation, i.e. for $c>0$ it acts in the same direction as the velocity. 
\subsubsection{Free motion}\label{fm}
For the free particle with dissipation (\ref{kgen}), the nonconservative action is 
\begin{eqnarray}
\Lambda(q_1,q_2,\dot q_1,\dot q_2)&=&\frac{m}{2}(\dot q_1^{2}-\dot q_2^{2})-\frac{c}{2}(q_{1}\dot q_{2}-q_{2}\dot q_{1})\nonumber\\
&=&2m\left[\dot{q}_{+}\dot{q}_{-}-\frac{c}{2m}\left(q_{-}\dot{q}_{+}-q_{+}\dot{q}_{-}\right)\right].
\end{eqnarray}
This action is invariant under time translations, under $SO(1,1)$ transformations (\ref{lorentz}), and
under translations it transforms by a total time derivative. It is also invariant under the PT transformation $(q_{+},q_{-},t)\rightarrow(q_{-},-q_{+},-t)$. 
The canonical momenta (\ref{p1}) and (\ref{p2}) are $p_1=m\dot{q}_1+\frac{c}{2}q_2$ and $p_2=m\dot{q}_2+\frac{c}{2}q_1$, 
and the Hamiltonian is (\ref{hamg}).
From Noether theorem there are four quantities, from (\ref{noncec3}) and (\ref{noncec4}) the energies $E_1$ and $E_2$, and from (\ref{jota2}) and (\ref{jota3}) the momenta $P_1$ and $P_2$, which are related to the Hamiltonian (\ref{ham}) and to the generator of translations (\ref{trans}). The conserved generator of $SO(1,1)$, (\ref{jota}), is
\begin{equation}
\tilde{\cal J}=q_1p_2-q_2p_1=2(q_{-}p_{+}-q_{+}p_{-}). \label{j1d}
\end{equation}
The $SO(1,1)$ invariance has the consequence that $E=\frac{1}{2}(E_{1}+E_{2})$ decomposes as $E=E_{+}+E_{-}$, where $E_\pm$ are given by (\ref{epm}). 
The equations of motion are $\ddot q_{\pm}\pm \frac{c}{m}\dot q_{\pm}=0$, with solutions $q_{\pm}(t)=\pm\frac{m}{c}v_{\pm}(0)(1-e^{\mp\frac{ct}{m}})+q_{\pm}(0)$. For $c>0$ the solution for $q_{-}$ is physically meaningless as its velocity and energy increase exponentially $\dot{q}_{-}(t)=v_{-}(0)e^{\frac{ct}{m}}$, unless the trivial solution is taken, in which case $E=\frac{m}{2}\dot{q}^2_{+}(0)e^{-\frac{2c}{m}t}$.
An application of this case is a particle constrained to move on a circle, with fixed radius $R$. The conservative Lagrangian is $L=\frac{mR^2}{2}\dot\theta^2$, and $K(\theta_1,\dot\theta_1,\theta_2,\dot\theta_2)=-\frac{cR^2}{2}(\theta_1\dot\theta_2-\theta_2\dot\theta_1)$.
\subsubsection{Free fall}
The nonconservative Lagrangian is
\begin{eqnarray}
\Lambda(q_1,q_2,\dot q_1,\dot q_2)&=&\frac{m}{2}(\dot q_1^{2}-\dot q_2^{2})+mg(q_1-q_2)-\frac{c}{2}(q_{1}\dot q_{2}-q_{2}\dot q_{1})\nonumber\\
&=&2m\left[\dot{q}_{+}\dot{q}_{-}+gq_{-}-\frac{c}{2m}\left(q_{-}\dot{q}_{+}-q_{+}\dot{q}_{-}\right)\right].
\end{eqnarray}
This action is invariant under time and position translations, the last up to a total time derivative. The Hamiltonian is
$H=\frac{2}{m}\left(p_{+}-\frac{c}{2}q_{+}\right)\left(p_{-}+\frac{c}{2}q_{-}\right)-2mgq_{-}$, and $E=E_{+}+E_{-}-mgq_{+}$, where $E_{\pm}$ are given by (\ref{epm}). The equations of motion are
\begin{eqnarray}
m\ddot{q}_{+}+2c\dot{q}_{+}+2mg&=&0,\label{eqmvf1}\\
m\ddot{q}_{-}-2c\dot{q}_{-}&=&0.\label{eqmvf2}
\end{eqnarray}
Thus, $q_{+}(t)=\frac{m}{c}[v_{+}(0)+\frac{mg}{c}](1-e^{-\frac{ct}{m}})-\frac{mg}{c}t+q_{+}(0)$. For $q_{-}(t)$ we get the free case of previous section, hence its consistent solution is $q_{-}(t)=0$.

\subsubsection{Damped oscillator}
The conservative Lagrangian is $L(q,\dot q)=\frac{m}{2}(\dot q^{2}-\omega^{2}q^{2})$ and the nonconservative potential is (\ref{kgen}). Thus
\begin{eqnarray}
\Lambda(q_1,q_2,\dot q_1,\dot q_2)&=&\frac{m}{2}(\dot q_1^{2}-\dot q_2^{2})-\omega^{2}(q_1^{2}-q_2^{2})-\frac{c}{2}(q_{1}\dot q_{2}-q_{2}\dot q_{1})\nonumber\\
&=&2m\left[\dot{q}_{+}\dot{q}_{-}-\omega^2q_{+}q_{-}-\frac{c}{2m}\left(q_{-}\dot{q}_{+}-q_{+}\dot{q}_{-}\right)\right].
\end{eqnarray}
This action is invariant under time translations, under $SO(1,1)$ transformations $\delta q_1=\eta q_2$ and $\delta q_2=\eta q_1$ \citep{mendes}, and under the PT transformation $(q_{+},q_{-},t)\rightarrow(q_{-},q_{-},-t)$. 
This Lagrangian has been proposed by Bateman \citep{bateman}. 
The  canonical momenta (\ref{p1}) and (\ref{p2}) are $p_1=m\dot{q}_1+\frac{c}{2}q_2$ and $p_2=m\dot{q}_2+\frac{c}{2}q_1$, and the Hamiltonian is
\begin{eqnarray}
H&=&\frac{1}{2m}\left(p_1^2-p_2^2\right)+\frac{m\omega_{-}^2}{2}\left(q_1^2-q_2^2\right)+\frac{c}{2m}\left(q_1p_2-q_2p_1\right)\nonumber\\
&=&\frac{2}{m}p_{+}p_{-}+2m\omega_{-}^2q_{+}q_{-}+\frac{c}{m}(p_{+}q_{-}-p_{-}q_{+}),\label{hamosc}
\end{eqnarray}
where $\omega_{-}^2=\omega^2-\frac{c^2}{4m^2}$. The conserved quantities which correspond to the two invariances of the nonconservative action are the Hamiltonian, and the generator of $SO(1,1)$ transformations $\tilde{\cal J}$ (\ref{j1d}). 
Further, the energies from the equations (\ref{noncec3}) and (\ref{noncec3}) are $E_1=\frac{m}{2}\dot q_1^2+\frac{m\omega^2}{2}q_1^2=\frac{1}{2m}p_1^2+\frac{m\omega^2}{2}q_1^2+\frac{c^2}{8m}q_2^2-\frac{c}{2m}q_2p_1$ and $E_2=\frac{m}{2}\dot q_2^2+\frac{m\omega^2}{2}q_2^2=\frac{1}{2m}p_2^2+\frac{m\omega^2}{2}q_2^2+\frac{c^2}{8m}q_1^2-\frac{c}{2m}q_1p_2$, from which can be obtained $H=E_1-E_2$, and $E=\frac{1}{2}(E_1+E_2)$ can be written as
\begin{equation}
E=\frac{1}{4m}\left(p_1^2+p_2^2\right)+\frac{m\omega_{+}^2}{4}\left(q_1^2+q_2^2\right)-\frac{c}{4m}\left(q_1p_2+q_2p_1\right),\label{enero}
\end{equation}
and satisfies $\frac{dE}{dt}=-2c\dot q_1\dot q_2$. As for the free case, due to the $SO(1,1)$ symmetry, in light cone coordinates (\ref{enero}) decomposes as $E=E_{+}+E_{-}$, where
\begin{equation}
E_{\pm}=\frac{1}{2m}p_{\pm}^2+\frac{m\omega_{+}^2}{2}q_{\pm}^2\mp\frac{c}{2m}p_{\pm}q_{\pm},\label{epmo}
\end{equation}
and $\omega_{+}^2=\omega_0^2+\frac{c^2}{2m^2}$. 

Computing the Poisson brackets (\ref{poisson}) of these quantities, it can be seen that there are four basic quantities $H$, $\tilde{\cal J}$, $E_{+}$ and $E_{-}$, which satisfy $\{H,\tilde{\cal J}\}=0$, $\{\tilde{\cal J},E_{\pm}\}=\pm 2E_{\pm}$, and 
\begin{equation}
\{E_{+},E_{-}\}=-\frac{1}{4}\left(\omega_{+}^2+\frac{c^2}{4m^2}\right)\tilde{\cal J}+\frac{c}{2m}\left(-\frac{1}{m}p_{+}p_{-}+m\omega_{+}^2x_{+}x_{-}\right).
\end{equation}
Actually, there is an $SO(1,2)$ algebra generated by $\tilde{\cal J}$ and $E_0^{\pm}=\frac{1}{2m}\left(p_{+}^2\pm p_{-}^2\right)+\frac{m\omega_{+}^2}{2}\left(q_{+}^2\pm q_{-}^2\right)$, i.e. $\{\tilde{\cal J},E_0^{\pm}\}=2E_0^{\mp}$ and $\{E_0^{+},E_0^{-}\}=\frac{1}{2}\omega_{+}^2\tilde{\cal J}$. If we make the substitution $\omega_{+}\rightarrow \omega_{-}$ in this algebra, we obtain the algebra of Feshbach and Tikochinsky \citep{feshbach}, see also \citep{chruscinski}.
The Hamiltonian can be decomposed as $H=H_0+\frac{c}{2m}\tilde{\cal J}$, where $H_0=\frac{2}{m}p_{+}p_{-}+2m\omega_{-}^2q_{+}q_{-}$ generates phase space $SO(2)$ transformations $\{H_0,q_{\pm}\}=-\frac{1}{m}p_{\pm}$, $\{H_0,p_{\pm}\}=m\omega_{-}^2q_{\pm}$. 

Note that the Hamiltonian and (\ref{enero}) satisfy $\{H,q_{\pm}\}=\frac{1}{m}\left(p_{\pm}\mp\frac{c}{2} q_\pm\right)$, $\{H,p_{\pm}\}=-\frac{1}{2}\left(\pm\frac{c}{m}p_\pm+2m\omega_{-}^2q_{\pm}\right)$, and $\{E,q_{\pm}\}=\frac{1}{2m}\left(p_{\pm}\mp\frac{c}{2} q_\pm\right)$, 
$\{E,p_{\pm}\}=-\frac{1}{2}\left(\pm\frac{c}{m}p_\pm+2m\omega_{-}^2q_{\pm}\right)$.

Further, in terms of expanding coordinates \citep{schuch2}, we write $q_{\pm}(t)=e^{\mp\frac{ct}{2m}}\rho_{\pm}$, and the equations of motion $\ddot q_{\pm}\pm \frac{c}{m}\dot q_{\pm}+\omega^2q_{\pm}=0$ become $\ddot\rho_{\pm}+\omega_{-}^2\rho_{\pm}=0$. Hence
\begin{equation}\label{sols}
\begin{array}{llr}
q_{\pm}(t)&=e^{\mp\frac{ct}{2m}}\left(A_\pm e^{i\omega_{-}t}+B_\pm e^{-i\omega_{-}t}\right) &\textnormal{if}\quad \omega^2>\frac{c^2}{4m^2},\\
q_{\pm}(t)&=e^{\mp\frac{ct}{2m}}(A_\pm+B_\pm t) &\textnormal{if}\quad\omega^2=\frac{c^2}{4m^2},\\
q_{\pm}(t)&=e^{\mp\frac{ct}{2m}}\left(A_\pm e^{\theta t}+B_\pm e^{-\theta t}\right) &\textnormal{if}\quad\omega^2<\frac{c^2}{4m^2},
\end{array}
\end{equation}
where $\theta^2=\frac{c^2}{2m^2}-\omega^2$. Thus $q_{+}(t)$ always describes a physical, decaying solution, unlike the case of $q_{-}(t)$, whose velocity increases exponentially.

\subsubsection{Central forces}
Consider two particles of masses $m_1$ and $m_2$, with position vectors $\vec x$ and $\vec y$, and which interact by a central potential $V(|\vec x-\vec y|)$.
A rotational invariant nonconservative potential, which corresponds to independent dissipative forces for these particles is
\begin{equation}
K(\vec x_1,\vec x_2,\vec y_1,\vec y_2,\dot{\vec x}_1,\dot{\vec x}_2,\dot{\vec y}_1,\dot{\vec y}_2)=-\frac{c_1}{2}(\vec x_1\dot{\vec x}_2-\vec x_2\dot{\vec x}_1)
-\frac{c_2}{2}(\vec y_1\dot{\vec y}_2-\vec y_2\dot{\vec y}_1).\label{kcf}
\end{equation}
In the center of mass coordinates $\vec r_1=\vec x_1-\vec y_1$, $\vec r_2=\vec x_2-\vec y_2$, $\vec R_1=\frac{1}{M}(m_1\vec x_1+m_2\vec y_1)$, and $\vec R_2=\frac{1}{M}(m_1\vec x_2+m_2\vec y_2)$, where $M=m_1+m_2$ is the total mass, (\ref{kcf}) becomes
\begin{eqnarray}
K=&-&\frac{c_1+c_2}{2}(\vec R_1\dot{\vec R}_2-\vec R_2\dot{\vec R}_1)-\frac{c_1m_2^2+c_2m_1^2}{2M^2}(\vec r_1\dot{\vec r}_2-\vec r_2\dot{\vec r}_1)\nonumber\\
&-&\frac{c_1m_2-c_2m_1}{2M}(\vec R_1\dot{\vec r}_2-\vec r_2\dot{\vec R}_1+\vec r_1\dot{\vec R}_2-\vec R_2\dot{\vec r}_1).
\end{eqnarray} 
Thus, if we set $c_1m_2-c_2m_1=0$, the center of mass decouples, and the nonconservative Lagrangian is $\Lambda=\Lambda_R+\Lambda_r$, where $\Lambda_R=\frac{M}{2}({\dot{\vec R}_1}^2-\dot{\vec R}_2^2)-\frac{c_1+c_2}{2}(\vec R_1\dot{\vec R}_2-\vec R_2\dot{\vec R}_1)$ represents the free particle of Section \ref{fm}, and $\Lambda_r$ corresponds to the case analyzed at the beginning of this section, with a central potential. The nonconservative Lagrangian is
\begin{eqnarray}
\Lambda_r&=&\frac{\mu}{2}(\dot{\vec r}_1^{\,2}-\dot{\vec r}_2^{\,2})-V(r_1)+V(r_2)-\frac{c}{2}(\vec r_1\dot{\vec r}_2-\vec r_2\dot{\vec r}_1),\nonumber\\
&=&2\mu\dot{\vec r}_{+}\dot{\vec r}_{-}-V(|\vec r_{+}+\vec r_{-}|)+V(|\vec r_{+}-\vec r_{-}|)-c\left(\vec r_{-}\dot{\vec r}_{+}-\vec r_{+}\dot{\vec r}_{-}\right),\label{eqmr}
\end{eqnarray}
where $\mu$ is the reduced mass, $c=\frac{c_1m_2}{M}$, and $\vec r_{\pm}=\frac{1}{2}(\vec r_1\pm \vec r_2)$. In fact, we could have considered (\ref{eqmr}) as a starting point for central forces.
The  canonical momenta are $\vec p_1=\mu\dot{\vec r}_1+\frac{c}{2}\vec r_2$ and $\vec p_2=\mu\dot{\vec r}_2+\frac{c}{2}\vec r_1$, and the Hamiltonian 
\begin{equation}
H=\frac{1}{2\mu}\left(\vec p_1^{\,2}-\vec p_2^{\,2}\right)+\frac{c}{2\mu}\left(\vec r_1\vec p_2-\vec r_2\vec p_1\right)+V(r_1)-V(r_2)-\frac{c^2}{8\mu}\left(r_1^2-r_2^2\right).\label{hamcf}
\end{equation}
The Lagrangian (\ref{eqmr}) is invariant under rotations, $\delta_\alpha (r_1)_i=\epsilon_{ijk}\alpha_j(r_1)_k$ and $\delta_\alpha (r_2)_i=\epsilon_{ijk}\alpha_j(r_2)_k$. From Noether theorem we get the energies $ 
E_1=\frac{1}{2\mu}\left(\vec p_1-\frac{c}{2}\vec r_2\right)^2+V(r_1)$, and $E_2=\frac{1}{2\mu}\left(\vec p_2-\frac{c}{2}\vec r_1\right)^2+V(r_2)$,
which satisfy $\frac{dE_1}{dt}=\frac{dE_2}{dt}=-c\dot{\vec r}_1\dot{\vec r}_2$. The energy (\ref{enero}) can be written as 
$E=E_{+}+E_{-}+\frac{1}{2}\left[V(r_1)+V(r_2)\right]$, where $E_{\pm}=\frac{1}{2m}\left(\vec{p}_{\pm}\mp\frac{c}{2} \vec{r}_{\pm}\right)^2$.
From (\ref{jota2}) and (\ref{jota3}) we get the angular momenta $\vec J_1=\mu(\vec r_1\times\dot{\vec r}_1)$ and $\vec J_2=\mu(\vec r_2\times\dot{\vec r}_2)$, which satisfy $\frac{d\vec J_1}{dt}=-c\mu(\vec r_1\times\dot{\vec r}_2)$ and $\frac{d\vec J_2}{dt}=c\mu(\vec r_2\times\dot{\vec r}_1)$. Thus, the angular momentum $\vec J=\frac{1}{2}\left(\vec J_1+\vec J_2\right)$ and the conserved generator of rotations $\vec{\cal J}=\vec J_1-\vec J_2$ are
\begin{eqnarray}
\vec J&=&\frac{1}{2}\left(\vec r_1\times\vec{p}_1+\vec r_2\times\vec{p}_2\right)=2(\vec{r}_{+}\times\vec{p}_{+}+\vec{r}_{-}\times\vec{p}_{-}),\label{jota3d1}\\
\vec{\cal J}&=&\vec{r}_1\times\vec{p}_1-\vec{r}_2\times\vec{p}_2=2(\vec{r}_{+}\times\vec{p}_{-}+\vec{r}_{-}\times\vec{p}_{+}).\label{jota3d2}
\end{eqnarray}
The angular momentum satisfies $\frac{d\vec J}{dt}=-\frac{c\mu}{2}(\vec r_1\times\dot{\vec r}_2-\vec r_2\times\dot{\vec r}_1)=-\frac{c}{m}(\vec r_{+}\times\vec{p}_{+}-\vec r_{-}\times\vec{p}_{-})$. We could have considered the rotations of $\vec r_1$ and $\vec r_2$ with independent parameters, i.e. $\delta (r_1)_i=\epsilon_{ijk}(\alpha_1)_j(r_1)_k$ and $\delta (r_2)_i=\epsilon_{ijk}(\alpha_2)_j(r_2)_k$. In this case the nonconservative potential $K$ is not invariant, and there are no conserved generators for these rotations.

The equations of motion are
\begin{equation}
\ddot{\vec r}_{\pm}\pm \frac{c}{\mu}\dot{\vec r}_{\pm}+\frac{1}{2\mu}\left\{\left[\frac{V'(r_1)}{r_1}\pm\frac{V'(r_2)}{r_2}\right]\vec r_{+}+\left[\frac{V'(r_1)}{r_1}\mp\frac{V'(r_2)}{r_2}\right]\vec r_{-}\right\}=0.\label{eqmc}
\end{equation}
Considering the unphysical dissipative force in the equation of $\vec r_{-}$, its consistent solution is $\vec r_{-}(t)=\vec 0$, from which follows $\vec r_1(t)=\vec r_2(t)$, and if we define $\vec r(t)=\vec r_{+}(t)$, then it satisfies
\begin{equation}
\ddot{\vec r}+\frac{c}{\mu}\dot{\vec r}+\frac{1}{\mu}\frac{V'(r)}{r}\vec r=0.\label{eqmcf}
\end{equation}
In this case the total angular momentum (\ref{jota3d1}) becomes $\vec J=2\vec{r}_{+}\times\vec{p}_{+}$, and satisfies $\frac{d\vec J}{dt}=-\frac{c}{2m}\vec J$, hence $\vec J(t)=\vec J_0 e^{-\frac{ct}{2\mu}}$ the motion taking place on a plane, with exponentially decreasing angular velocity. 

\subsection{Non linear dissipation}
Consider a particle with a nonlinear nonconservative potential 
\begin{equation}
K=-q_{-}\kappa(q_{+},\dot{q}_{+})\dot{q}_{+}, 
\end{equation}
where 
\begin{equation}
\kappa(q_{+},\dot{q}_{+})=c_1(q_{+})+c_2(q_{+})|\dot{q}_{+}|+\cdots+c_n(q_{+})|\dot{q}_{+}|^{n-1}.
\end{equation}
The nonconservative Lagrangian is
\begin{eqnarray}
\Lambda(q_1,q_2,\dot q_1,\dot q_2)
=2m\dot{q}_{+}\dot{q}_{-}-V(q_{+}+q_{-})+V(q_{+}-q_{-})-q_{-}\kappa(q_{+},\dot{q}_{+})\dot{q}_{+}.\label{lnl}
\end{eqnarray}
The momenta are $p_{+}=m\dot{q}_{+}$, and $p_{-}=m\dot{q}_{-}-\frac{1}{2}q_{-}(c_1+2c_2|\dot{q}_{+}|^2\cdots+nc_n|\dot{q}_{+}|^{n-1})$. Thus $\dot{q}_{+}=\frac{1}{m}p_{+}$, and $\dot{q}_{-}=\frac{1}{m}p_{-}+\frac{1}{2m}q_{-}\left(c_1+\frac{2c_2}{m}|p_{+}|+\cdots+\frac{nc_n}{m^{n-1}}|p_{+}|^{n-1}\right)$, hence the Lagrangian is regular.
For a free particle, i.e. with vanishing conservative potential $V$, the Lagrangian is invariant under $q_{+}$ translations. 

The equations of motion are
\begin{equation}
\ddot{q}_{+}+\frac{1}{2m}\left(c_1+c_2|\dot{q}_{+}|+\cdots+c_n|\dot{q}_{+}|^{n-1}\right)\dot{q}_{+}=0,\label{eqmnl1}
\end{equation}
and for constant $c_1,\dots,c_n$ 
\begin{eqnarray}
&&\ddot{q}_{-}-\frac{1}{2m}\left(c_1+2c_2|\dot{q}_{+}|+\cdots+nc_n|\dot{q}_{+}|^{n-1}\right)\dot{q}_{-}\nonumber\\
&&-\frac{1}{2m}\left[2c_2+6c_3|\dot{q}_{+}|\cdots+n(n-1)c_n|\dot{q}_{+}|^{n-2}\right]\frac{|\dot{q}_{+}|}{\dot{q}_{+}}\ddot{q}_{+}q_{-}=0.\label{eqmnl2}
\end{eqnarray}
Thus, the equation of $q_{+}$ decouples. Otherwise, considering that the velocity $\dot q_{+}(t)$ and the acceleration $\ddot q_{+}(t)$ tend to zero due to the dissipation, the interaction terms in (\ref{eqmnl2}) can be treated perturbatively with respect to the term $-\frac{c_1}{2m}\dot{q}_{-}$. In this case, the zeroth order solution for $q_{-}(t)$ is the free particle with linear dissipation, hence the trivial solution must be considered for it as previously shown, and in consequence  the perturbed solution will be also the trivial one. 
For $n=2$, with $c_1$ and $c_2$ constant, the Hamiltonian and the energy are
\begin{eqnarray}
H&=&\frac{2}{m}p_{+}p_{-}+\frac{1}{m}q_{-}\left(c_1+\frac{c_2}{m}p_{+}\right)p_{+},\\
E&=&\frac{1}{2m}p_{+}^2+\frac{1}{m}\left[p_{-}+\frac{1}{2}q_{-}\left(c_1+\frac{2c_2}{m}|p_{+}|^2\right)\right]^2.
\end{eqnarray}
The equations of motion are 
\begin{eqnarray}
\ddot{q}_{+}+\frac{c_1}{2m}\dot{q}_{+}+\frac{c_2}{2m}|\dot{q}_{+}|\dot{q}_{+}=0,\\
\ddot{q}_{-}-\frac{c_1}{2m}\left(1+2\frac{c_2}{c_1}|\dot{q}_{+}|\right)\dot{q}_{-}-\frac{c_2}{m}\frac{|\dot{q}_{+}|}{\dot{q}_{+}}\ddot{q}_{+}q_{-}=0, 
\end{eqnarray}
with solution for $q_{+}$
\begin{equation}
q_{+}(t)=q_{+}(0)+\frac{2m}{c_2}\log\left[1+\frac{c_2}{c_1}v_{+}(0)\left(1-e^{-\frac{c_1t}{2m}}\right)\right],\label{qnl1}
\end{equation}
and for $q_{-}$
\begin{eqnarray}
{q}_{-}(t)&=&\left[1+\frac{c_2}{c_1}v_{+}(0)\left(1-e^{-\frac{c_1t}{2m}}\right)\right]\Bigg\{q_{-}(0)\nonumber\\
&&+\frac{1}{c_1}\left[c_2v_{+}(0)q_{-}(0)-2mv_{-}(0)\right]\left(1-e^{\frac{c_1t}{2m}}\right)\Bigg\},\label{qnlm1}
\end{eqnarray}
which is unphysical, unless the factor of $e^{\frac{c_1t}{2m}}$ vanishes, i.e. the initial velocity of $q_{-}$ is related to its initial position by $v_{-}(0)=\frac{c_2}{2m}v_{+}(0)q_{-}(0)$.  However, this solution is proportional to $q_{-}(0)$, i.e. if the scale of this parameter is set so that it is at the null point $q_{-}(0)=0$, we have the trivial solution. In order to avoid this meaningless behavior, which is due to the lack of invariance of (\ref{lnl}) under $q_{-}$ translations, the trivial solution must be considered, as in the previous cases.

In the case of purely quadratic forces, i.e. $c_1=0$, (\ref{qnl1}) and (\ref{qnlm1}) become
\begin{equation}
q_{+}(t)=q_{+}(0)+\frac{2m}{c_2}\log\left[1+\frac{c_2}{2m}v_{+}(0)t\right].\label{qnl2}
\end{equation}
and a uniformly accelerated motion for $q_{-}$
\begin{equation}
{q}_{-}(t)=q_{-}(0)+v_{-}(0)t-\frac{c_2}{4m^2}\left[c_2v_{+}(0)q_{-}(0)-2mv_{-}(0)\right]v_{+}(0)t^2.\label{qnlma1}
\end{equation}
This solution is unphysical, unless the velocity $v_{-}(0)$ satisfies the same condition as in the preceding case, with the same shortcoming of being proportional to $q_{-}(0)$.

If a constant force (free fall) is added to the previous case, the nonconservative Lagrangian becomes
\begin{eqnarray}
\Lambda(q_1,q_2,\dot q_1,\dot q_2)
=2m\dot{q}_{+}\dot{q}_{-}+2mgq_{-}-q_{-}\kappa(q_{+},\dot{q}_{+})\dot{q}_{+}. 
\end{eqnarray}
Thus, only the equation for $q_{+}$ is modified $\ddot{q}_{+}-2mg+\frac{c_1}{2m}\dot{q}_{+}+\frac{c_2}{2m}|\dot{q}_{+}|\dot{q}_{+}=0$, with solution
\begin{equation}
q_{+}(t)=a_1-\frac{c_1 t}{2 c_2}+\frac{2m}{c_2}\log \left\{\cosh \left[\frac{c_1}{2}\sqrt{1+8\frac{mg c_2}{c_1}}\left(a_2+\frac{1}{2m}t\right) \right]\right\},
\end{equation}
where $a_1$ and $a_2$ are integration constants. Consistently, the velocity of the particle becomes constant after a while. In this case the equation for $q_{-}$ cannot be solved analytically but a numerical inspection shows that in general is has an unphysical behavior. 

\section{Conclusions}
We have studied the doubled variable Lagrange formulation for systems that are not conservative due the action of dissipative forces, that can be modelated by a nonconservative potential, bearing in mind two aspects relevant for quantization as follows. First, Noether theorem as the association of conservation laws to symmetries of the system, considering that in this case there are symmetries of the original conservative system, as well as symmetries of the nonconservative potential. Secondly, the consideration of all the degrees of freedom as contributing to the physical dynamics of the system. Thus, there are conserved quantities that generate the exact symmetries of the nonconservative Lagrangian, e.g. the Hamiltonian which generates time translations, as well as nonconserved quantities, which are conserved in the absence of the dissipative forces, energy, angular momentum, etc. The nonconserved quantities have the same expressions in configuration space as in the conservative theory, and appear in doubled versions, corresponding to each type of variables. From the equations satisfied by the nonconserved quantities follow the conservation equations of the symmetry generators of the nonconservative system, in particular the Hamiltonian. Additionally there are conserved quantities for the symmetries that mix both types of variables and which have no correspondence for the conservative system. 
In the example of the damped harmonic oscillator, the conserved current of the $SO(1,1)$ symmetry of the action is one of the generators of the $SO(1,2)$ algebra of Feshbach and Tikochinsky \citep{feshbach}, and the other two generators are related to the energies (\ref{epmo}).
Regarding the second aspect considered in this paper, the two sectors of the system appear to divide in one physical sector, with physically consistent solutions, and another one which has unphysical solutions, i.e. solutions whose energy increases steadily. In the absence of external forces, which would contribute to the equations of motion by inhomogeneous terms, this second sector has always the trivial solution. Moreover, at least for nonlinear dissipative forces, under specific conditions, this sector ostensibly contains physically meaningful solutions, which however are shown to be inconsistent.
We considered various examples, and analyzed them from the point of view of this paper. For most of them there are analytical solutions in both sectors. For linear dissipation we considered the free motion, the free fall, the harmonic oscillator, central forces, and for nonlinear dissipation the free motion. Regarding central forces in three dimensions, we considered two particles with a rotational invariant nonconservative Lagrangian. It turns out that if the dissipation constants and masses are suitably related, the center of mass decouples from the relative motion, the last having a nonconservative Lagrangian with a central potential and a dissipative force. Further, taking the trivial solution for the unphysical sector leads to the usual picture, with the direction of rotation conserved, and a damped angular momentum magnitude. For nonlinear dissipation we considered a somewhat general case, formulated in such a way that the equation for the physical sector decouples from the unphysical sector, and for the quadratic case we considered a free particle and a particle under free fall.

\vspace{.5cm}
{\bf Acknowledgements}

We thank BUAP Grant RARC-EXC16-G, and PFCE-SEP for the support. We thank also C.A. Jano, I. L\'opez and G. Cruz for useful discussions.

\end{document}